\pdfoutput=1
\documentclass[letterpaper]{jacow}
\makeatletter%
	\ifboolexpr{bool{xetex}}
	 {\renewcommand{\Gin@extensions}{.pdf,%
	                    .png,.jpg,.bmp,.pict,.tif,.psd,.mac,.sga,.tga,.gif,%
	                    .eps,.ps,%
	                    }}{}
\makeatother

\ifboolexpr{bool{xetex} or bool{luatex}} 
 {}                                      
 {\usepackage[utf8]{inputenc}}           

\usepackage[USenglish]{babel}			 

\usepackage[final]{pdfpages}
\usepackage{multirow}
\usepackage{ragged2e}
\usepackage{pbox}
\usepackage{lineno}
\usepackage{amsmath}

\ifboolexpr{bool{jacowbiblatex}}%
 {%
  \addbibresource{jacow-test.bib}
  \addbibresource{biblatex-examples.bib}
 }{}
\begin{document}

\title{A FODO racetrack ring for \lowercase{nu}STORM: design and optimization\thanks{Work supported by Fermilab, Operated by Fermi Research Alliance, LLC under Contract No. DE-AC02-07CH11359 with the United States Department of Energy.}}

\author{A. Liu\thanks{aoliu@fnal.gov}, A. Bross, D. Neuffer, Fermi National Accelerator Laboratory, Batavia, USA}

\maketitle

\begin{abstract}
The goal of nuSTORM is to provide well-defined neutrino beams for precise measurements of neutrino cross-sections and oscillations. The nuSTORM decay ring is a compact racetrack storage ring with a circumference of $\sim$480 m that incorporates large aperture (60 cm diameter) magnets. There are many challenges in the design. In order to incorporate the Orbit Combination Section (OCS), used for injecting the pion beam into the ring, a dispersion suppressor is needed adjacent to the OCS. Concurrently, in order to maximize the number of useful muon decays, strong bending dipoles are needed in the arcs to minimize the arc length. These dipoles create strong chromatic effects, which need to be corrected by nonlinear sextupole elements in the ring. In this paper, a FODO racetrack ring design and its optimization using sextupolar fields via both a Genetic Algorithm (GA) and a Simulated Annealing (SA) algorithm will be discussed.
\end{abstract}

\section{INTRODUCTION}
\begin{figure*}[!h]
  \centering
  \includegraphics[width=0.6\textwidth]{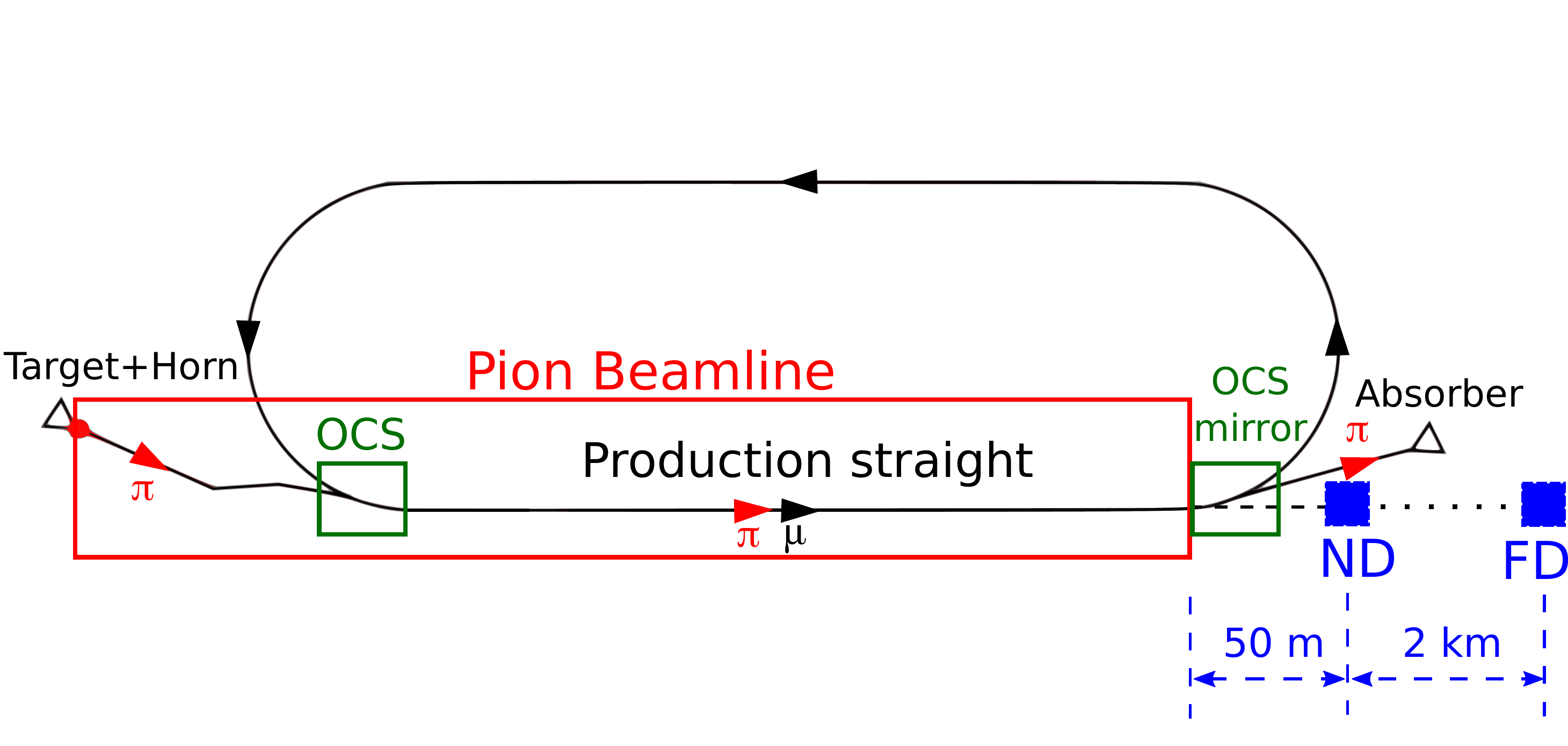}
  \caption{The schematic of the beamline structure in the nuSTORM facility}
  \label{Figure_Layout}
\end{figure*}

The schematic drawing of the nuSTORM \cite{nuSTORM_Proposal, nuSTORM_PRD} facility is presented in Figure~\ref{Figure_Layout}.  A high-intensity proton beam of 120 GeV is directed onto a target, producing a large spectrum of secondary pions within a wide range of momenta. The forward pions centered around 5 GeV/c are focused by a collection magnetic horn and inserted through a chicane into the straight section of the storage ring. Muons produced within the storage ring acceptance by pion decay along the straight section are stored for approximately the muon lifetime before they are dumped prior to the next spill \cite{NeufferStochastic}. In turn muons decay producing neutrino beams of known flux and flavor ($\mu^+\rightarrow e^++\bar{\nu}_{\mu}+\nu_e$ or $\mu^-\rightarrow e^-+\bar{\nu}_e+\nu_\mu$). Considering that a neutrino experiment does not require the beam to be bunched, and the fact that synchrotron radiation loss of the muons can be neglected, the nuSTORM ring does not require RF cavities.

\begin{figure}[!h]
	\centering
	\includegraphics[width=0.48\textwidth]{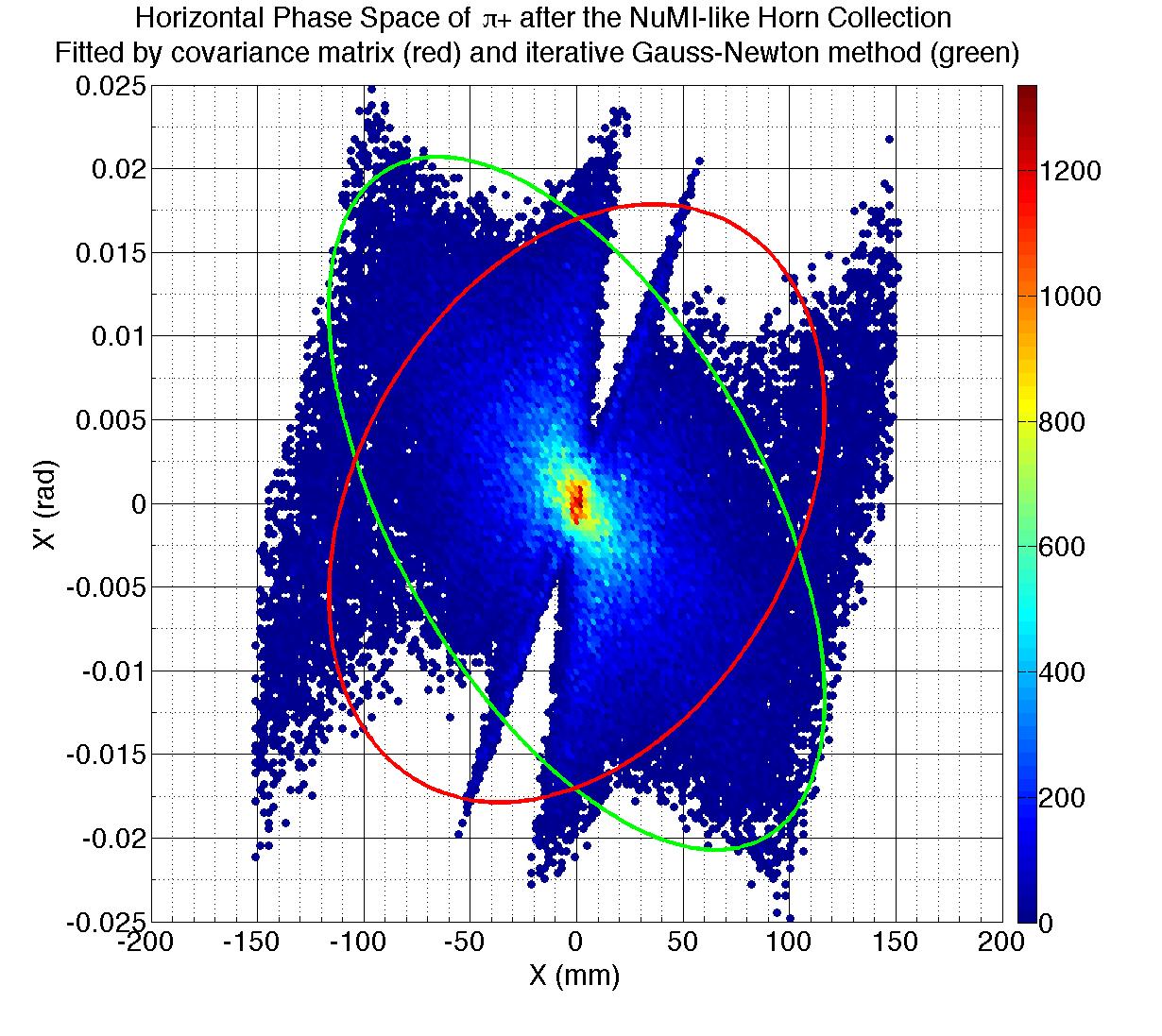}
	\caption{The distribution of $5\pm1$ GeV/c pions in the transverse phase space, as generated using MARS15\cite{MARS15}, and the corresponding fitted 2 $\mu$m$\cdot$rad Gaussian acceptance ellipse. The fitting results from the IGN method and the covariance matrix are shown in green and red, respectively. The IGN method conserves the core shape of the distribution.}
	\label{Figure_AfterHorn}
\end{figure}

The critical design challenge is in storing the maximal number of muons for enough number of turns. The nuSTORM solution is to design a large-acceptance storage ring (accepting a momentum spread of $\delta=dP/P_0\approx\pm10\%$ and an unnormalized transverse RMS emittance 300 $\mu$m$\cdot$rad) and to inject higher-momentum pions from a large-acceptance pion beamline (PBL) into the production straight section. The PBL uses stochastic injection to guide the pion beam into the 3.8 GeV/c decay ring \cite{Ao_iPAC2013, Ao_iPAC2013_talk, NIMA_pion_beamline}, for which the center momentum is chosen to optimize the neutrino flux in the desired energy range. The stochastic injection is implemented in a Orbit Combination Section (OCS) designed to combine the orbits of the 5 GeV/c reference pion and 3.8 GeV/c reference muon. The OCS creates a large dispersion at the entrance of the section to set the orbit of the 5 GeV/c pion $\sim$ 25 cm away from that of the 3.8 GeV/c muon. After the OCS, the dispersion is suppressed to zero such that the orbits are combined. The performance of the PBL design was investigated in G4Beamline\cite{G4BL_website}. Figure~\ref{Figure_AfterHorn} shows the distribution of pions in the transverse phase space at the downstream face of the magnetic collection horn. An ellipse showing the fitted 300 $\mu$m$\cdot$rad acceptance (6 times the RMS emittance) aperture is displayed. An Iterative Gauss-Newton (IGN) fitting method was used to remove the fitting bias caused by particles with extra-large emittance. Based on the PBL design, the horn can be optimized to obtain a $15\%$ increase in neutrino flux \cite{NIMA_Ao_Horn}.

The goal of the nuSTORM ring is to circulate the muon beam generated in the production straight section, as described above. The design of an OCS type stochastic injection requires the dispersion $D_x$ to be large at the injection point, so that the two separate orbits can be combined. However, a large $D_x$ naturally enlarges the beam size, and is hence harmful to the transmission in the ring. Accordingly, directly next to the injection point, $D_x$ needs to be at a much lower value. In fact, $D_x$ should be suppressed anywhere except at the OCS or the OCS mirror, which directs the pions that remain at the end of the production straight to an absorber. This needs strong dipoles and quadrupoles, which create large natural chromaticities. Sextupoles, placed where the linear dispersion $D_x$ is large for correcting the natural chromaticity, also introduce higher order dispersion terms, which for large $\delta$ are non-negligible and enlarge the beam size. In addition, sextupoles introduce non-linear resonances and geometric aberrations. For a decay ring long dispersion-free straight sections are desired, and the ratio of straight to arc needs to be as large as possible. 

As a consequence, inserting sextupoles in the nuSTORM FODO ring is challenging. An algorithm is therefore needed to choose the best sextupole correction scheme to optimize the acceptance of the large muon beam, directly from multi-particle tracking, rather than trying to correct all the chromatic and geometric aberration terms. A Genetic Algorithm (GA) or a Simulated Annealing (SA), both of which are metaheuristic, can be applied for this purpose. In this paper, the linear optics design of the nuSTORM FODO ring and the sextupole correction are discussed.

\section{Linear optics design}
\begin{figure*}[!h]
	\centering
	\includegraphics[width=0.8\textwidth]{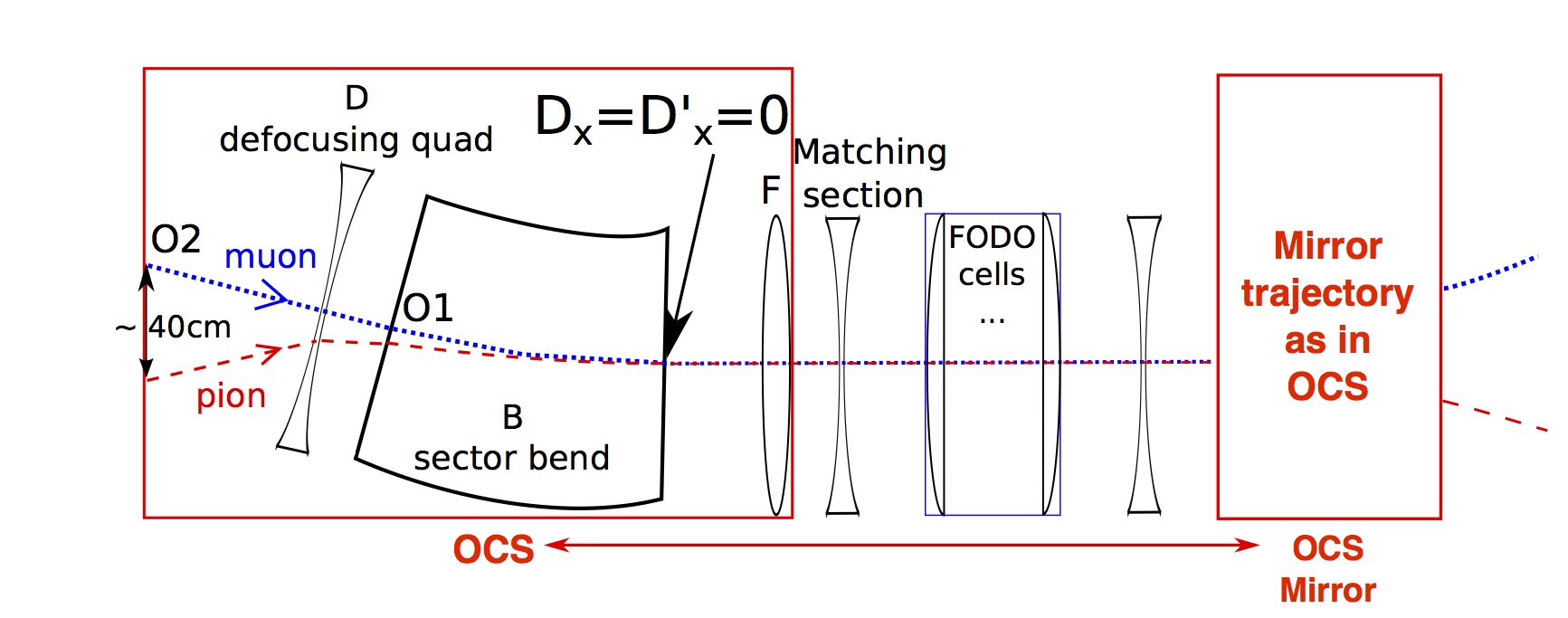}
	\caption{The schematic drawing of the OCS and production straight part of the decay ring.}
	\label{Figure_ring_schem1}
\end{figure*}
\begin{figure}[!h]
	\centering
	\includegraphics[width=0.4\textwidth]{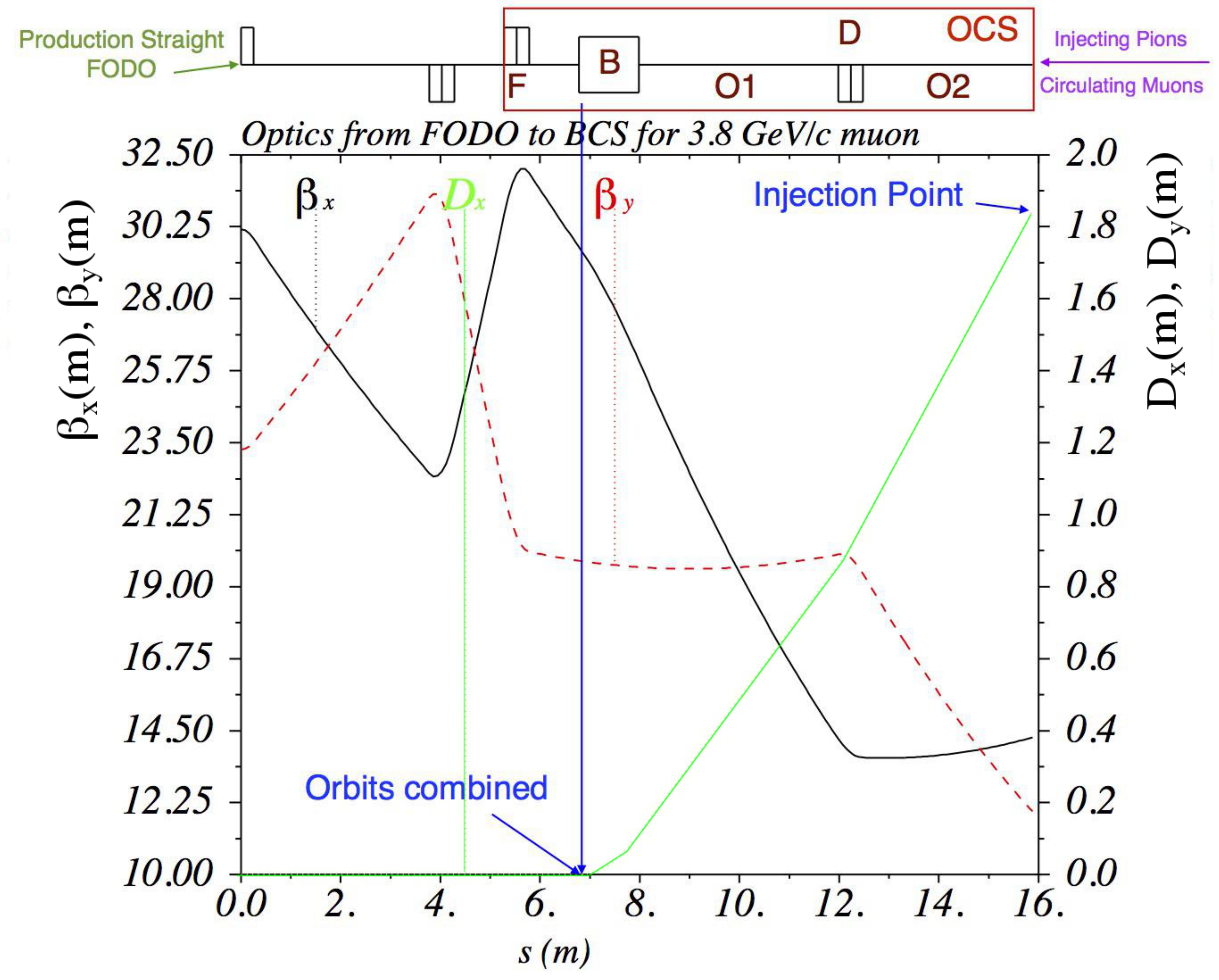}
	\caption{The 3.8 GeV/c reference $\mu$ optics from the FODO straight to the injection point. Both circulating muons and injected pions travel from right to left.}
	\label{Figure_FODO2OCSmu}
\end{figure}
The nuSTORM decay ring incorporates the OCS described above. A schematic drawing of the OCS and production straight section of the ring is shown in Figure~\ref{Figure_ring_schem1}. The injection happens at the leftmost of the figure, while the undecayed pions can be extracted in the ``mirror OCS" at the end of the production straight as mentioned above.

The production straight is composed of standard FODO cells, within which muons decay to useful neutrinos. The FODO cells are designed to accommodate both muons and pions by providing separate periodic optics for the two reference momenta. The optics of the 3.8 GeV/c reference $\mu$ from the FODO cell to the injection point is shown in Figure~\ref{Figure_FODO2OCSmu}, in which beams move from right to left. $D^{(0)}_x$ at the injection point was intentionally designed to be large enough to allow for a 40 cm separation from the two reference orbits at injection. 

There are two schemes for the arc design. In the first scheme, the average linear dispersion is kept small by implementing an  achromat cell structure. This scheme was ruled out in the early stage of the design because of the difficulty in obtaining a small natural chromaticity and in implementing sextupolar field correction in the non-dispersive sections. The other scheme uses combined-function dipoles with nonzero $D^{(0)}_x$ in arcs. Two options for this scheme are discussed in this article. The first option uses arcs with superconducting magnets, while the second uses normal conducting arcs. 

Despite the resulting large beam size, large $\beta_{\perp}$ are preferable in the production straight section because of the smaller divergence of the produced neutrino beam. For the opposite straight section, where the neutrino beam is not used and which is used for adjusting the total tunes, a smaller $\beta_{\perp}$ has been chosen. As a result, the ring has only one reflection symmetry, for which the reflection axis is the connection between the straight section centers. 

The circumference of the ring is approximately 480 meters, which is slightly bigger than the Fermilab Booster. With arcs of approximately 60 meters as in the preferred design scheme being discussed later, the fraction of the ring circumference that is useful for neutrino production is $\sim0.39$. This geometry also makes it possible to utilize the intense $\nu_\mu$ or $\bar{\nu}_\mu$ beam from the injected $\pi^+$ or $\pi^-$ decay, as the neutrino signals can be distinguished from electron and muon neutrinos from muon decay through timing. 

With the lifetime of 3.8 GeV/c muons in the lab frame being $\sim79\mu$s, approximately 87\% of them will decay in 100 turns. Consequently, the number of muons that survive 100 turns (with decay turned off in G4Beamline) in tracking is used as the ``benchmark" for comparing the ring designs. In this section, two arc designs, one with superconducting combined function magnets and the other with normal-conducting combined function magnets, are presented. The superconducting and normal-conducting magnets allow the pole-tip field to go up to 4 T and 2 T, respectively.

\subsection{Arc design with superconducting combined function magnets}
\begin{figure}[!h]
  \centering
  \includegraphics[width=0.4\textwidth]{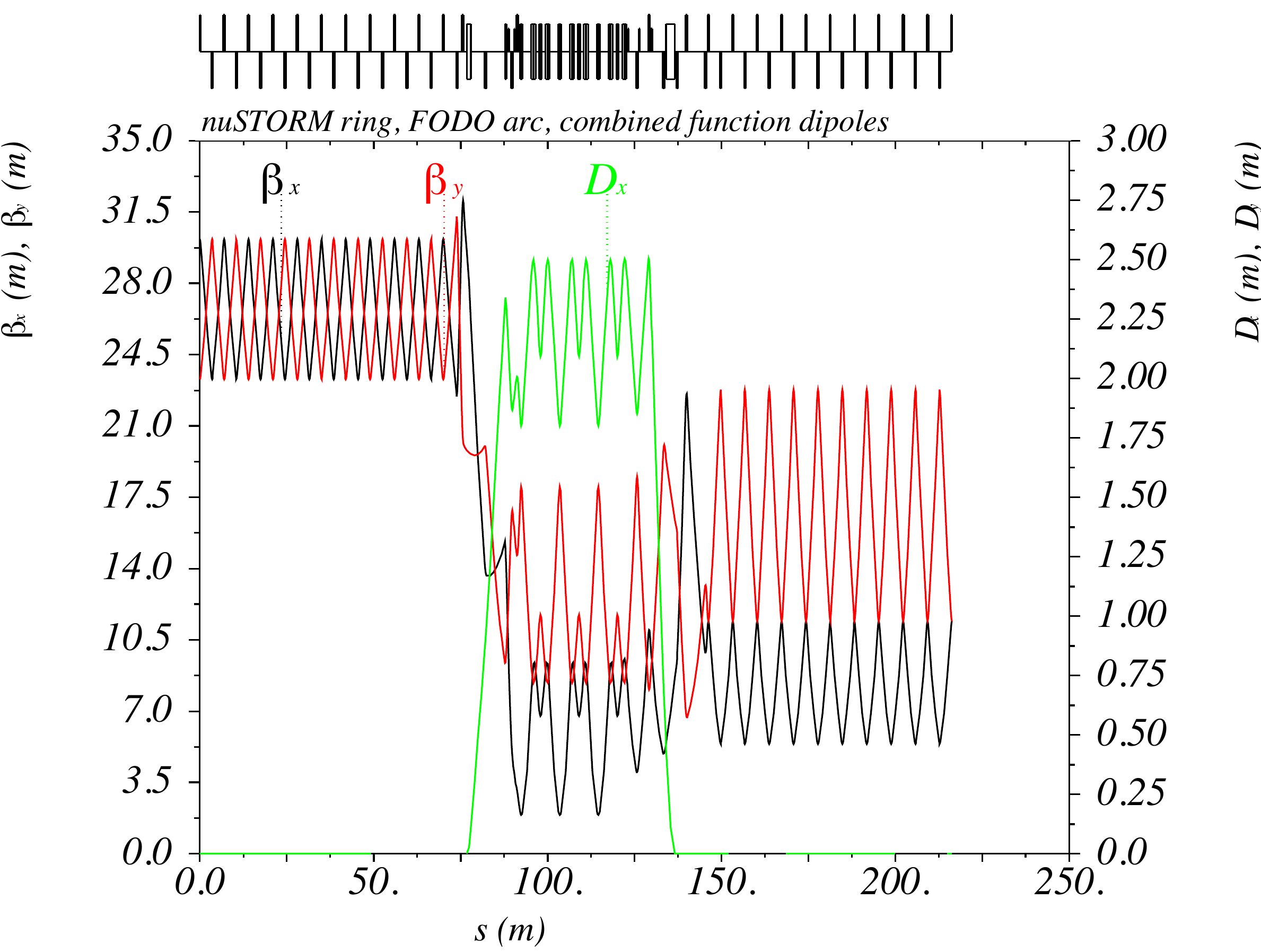}
  \caption{The Twiss functions of the ring design with arcs formed by superconducting combined-function magnets.}
  \label{Figure_SC_DKring}
\end{figure}

In this design, superconducting combined-function magnets were used. The combined-function magnets have the advantage of making a compact arc, and reducing the natural chromaticity, $C_x$, which is given by \cite{SY_book}
\begin{equation*}
\begin{aligned}
	C_x = \frac{1}{4\pi}\oint\beta_x \Big{[}&-\frac{2}{\rho^2}+K+\frac{D_x}{\rho}\left(\frac{1}{\rho^2}-3K\right)-\left(\frac{1}{\rho}\right)'{D_x}'+ \\
	&\frac{\gamma_x}{\beta_x\rho}D_x\Big{]} ds
\end{aligned}
\end{equation*}
where the combined function magnets, although with a strong dipole field and therefore smaller bending radius, suppress $C_x$ by adding the counter-part $K$. The vertical chromaticity is not arbitrarily suppressed by introducing the combined function magnets. The linear optics of this design are shown in Figure~\ref{Figure_SC_DKring}. The key parameters of the ring are shown in Table~\ref{Table_concept1}.
\begin{table}[!h]
	\centering
	\caption{Parameters of the decay ring design with superconducting combined function dipole arcs}
	\begin{tabular}{|p{0.25\textwidth}|p{0.2\textwidth}|}
		\hline
		Parameters & Values (units)\\
		\hline
		\hline
		Central momentum $P_{0,\mu}$ & 3.8 (GeV/c)\\
		Circumference & 488.5 (m)\\
		Arc length & 59.8 (m)\\
		Production straight L & 181.56 (m)\\
		Non-production straight L & 187.34 (m)\\
		($\nu_x$, $\nu_y$) & (8.12, 4.63)\\
		($C_x=d\nu_x$/$d\delta$, $C_y=d\nu_y$/$d\delta$) & (-4.29,-6.62)\\
		\hline
	\end{tabular}
	\label{Table_concept1}
\end{table}

\begin{figure*}[htp]
	\centering
	\includegraphics[width=0.35\textwidth]{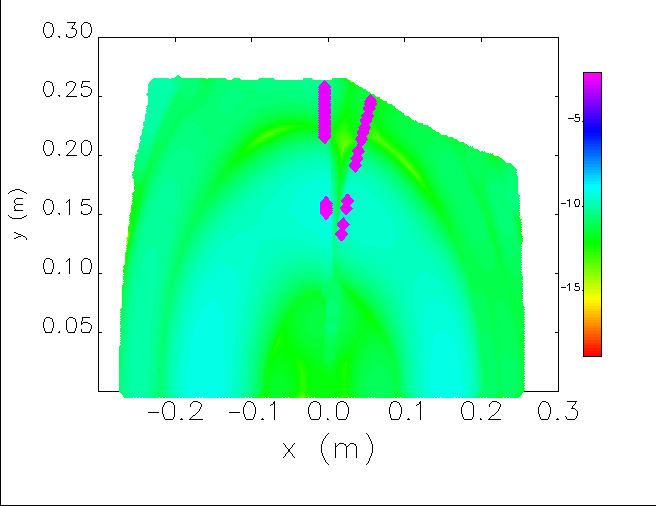}
	\includegraphics[width=0.35\textwidth]{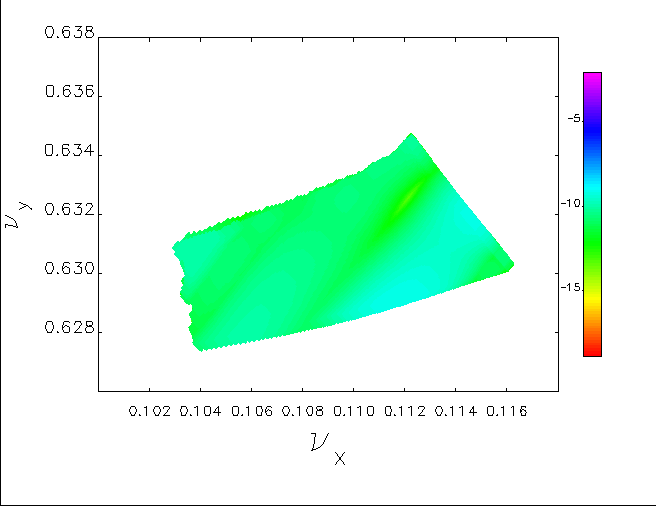}\\
	\includegraphics[width=0.35\textwidth]{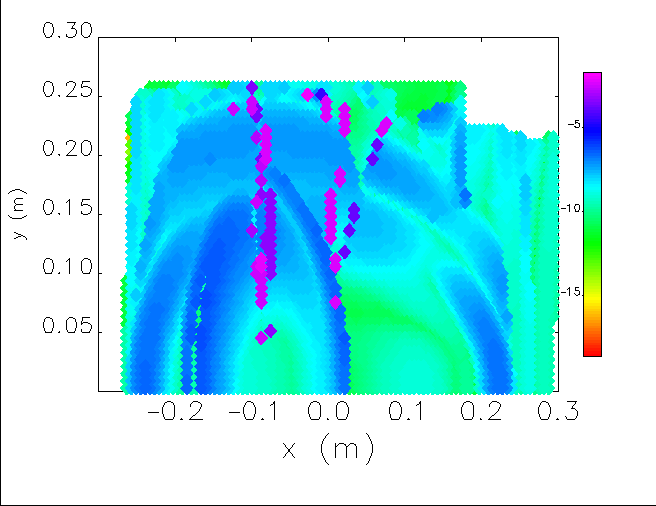}
	\includegraphics[width=0.35\textwidth]{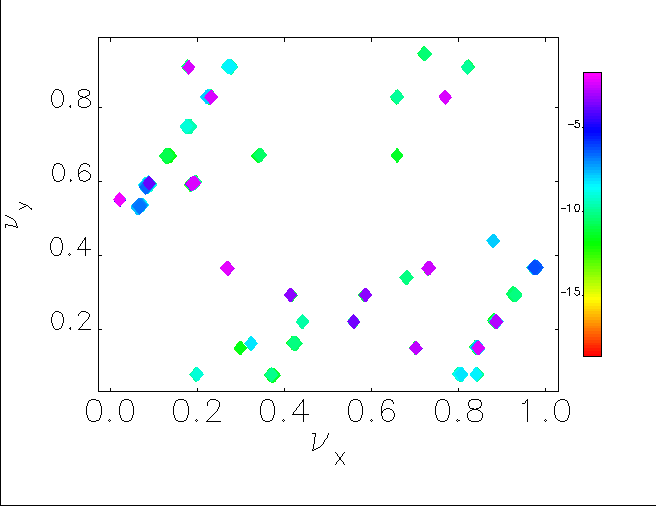}
	\caption{The Frequency Map Analysis (FMA) of the ring design with superconducting combined function dipoles in the arcs in the X-Y space (left) and on the betatron tune plane (right), for on-momentum particles (upper) and off-momentum particles within the full range of $\delta=\pm0.1$ (lower) are analyzed. No sextupole corrections were added.}
	\label{Figure_FMA_concept1}
\end{figure*}

However, as explained above, the natural linear chromaticity is only one of the many factors that determines the ring performance. In order to find the tune footprint of the ring lattice, a Frequency Map Analysis (FMA) method, analyzed by $elegant$ \cite{elegant_MB}, is employed to gain more insight, especially on the nonlinear properties of the ring. First the tune shifts with amplitude are checked with on-momentum particles. The FMA of on-momentum particles with proper aperture limits added in $elegant$ is shown in the upper plots of Figure~\ref{Figure_FMA_concept1}. The lower plots show that the chromatic effects of this design are investigated by doing the FMA analysis on particles within a full momentum range of $\delta\in\pm10\%$.

As illustrated by the plots, the natural chromaticities are greatly reduced in this scenario. As in an ordinary FODO ring design the chromaticities are slightly larger than the betatron tunes of the ring \cite{SY_book}. The tune dependence on the particle action is small for this design, however the chromatic driving terms still dilute the tune distribution in the tune plane without a clear pattern. Because of the combination of chromatic and geometric nonlinear effects in the ring, the FMA yielded scattered accepted tunes on the tune plane. Thus the acceptance of the lattice is better determined by multi=particle tracking using for instance the symplectic tracking of MADX-PTC \cite{MADX, MADX_PTC_Schmidt}. The tracking was done for a muon beam with Gaussian transverse distribution, for which the transverse admittance is 2000 $\mu$m$\cdot$rad, and with a uniform momentum distribution within $3.8\pm10\%$ GeV/c. The data correspond to $10^5$ particles, if not specified otherwise. Apertures for each magnet element are defined such that particles outside the apertures are intercepted and lost. Without the sextupole correction, the number of particles surviving 100 turns is 58\%, and around $\sim10\%$ of the starting particles are lost within the first turn. More specifically, the loss is highest at both the OCS and the OCS mirror, where both $\beta_{x}$ and $D_x$ are large. This large loss is to some extent inevitable because of the stochastic injection functionality of the OCS.

\subsection{Arc design with normal-conducting combined function magnets}
\begin{figure}[htp]
	\includegraphics[width=0.45\textwidth]{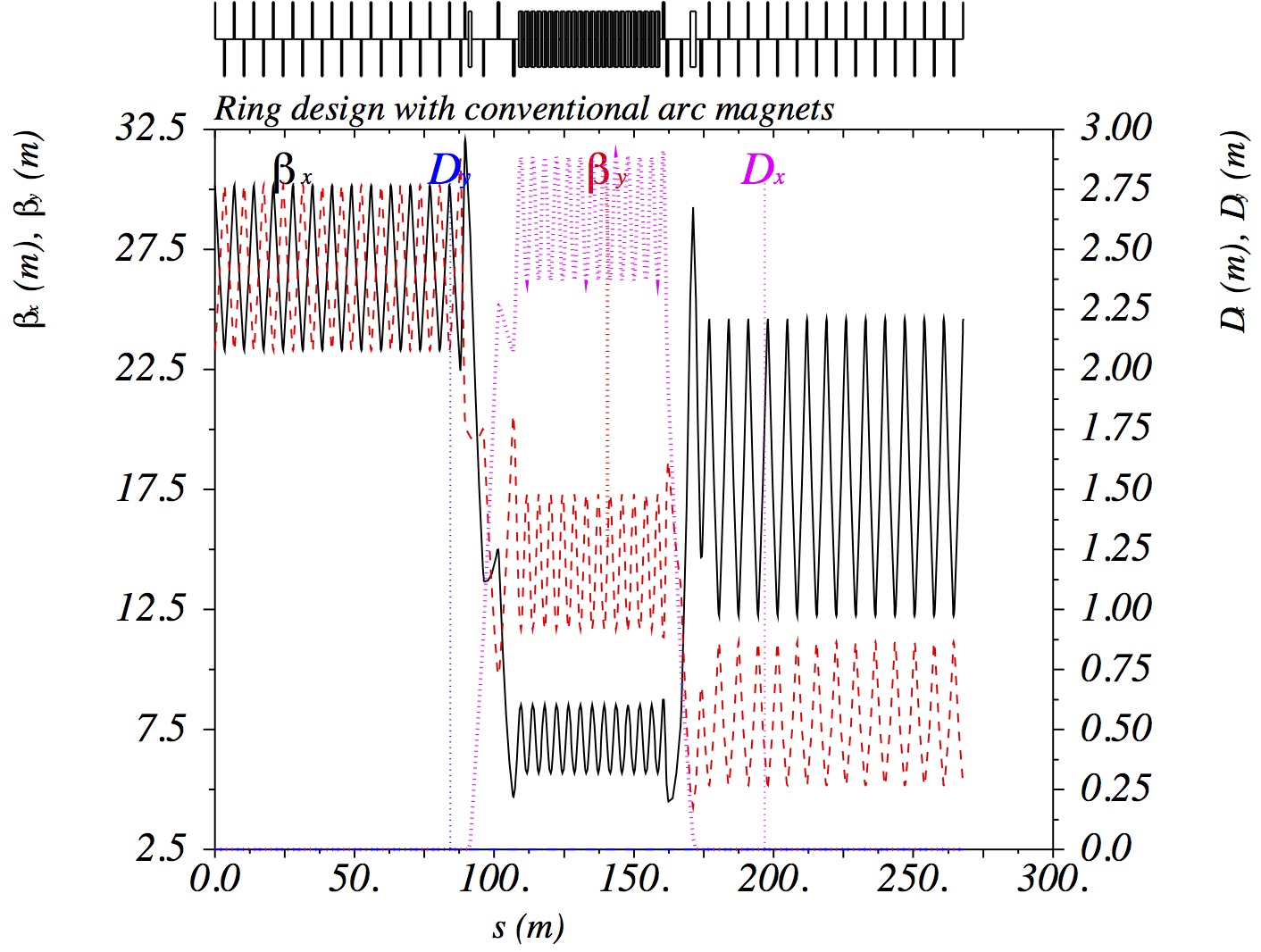}
	\caption{The linear optics of a nuSTORM decay ring design with conventional combined function dipoles in the arcs. $\beta_x$: black, $\beta_y$: red, $D_x$: green.}
	\label{Figure_Twiss_concept3}
	\clearpage
\end{figure}

Another possible design option was proposed to avoid the complexity of introducing cryogenic equipment into the facility. The arc was redesigned to include only normal-conducting combined-function dipoles. The parameters of the new design option are listed in Table \ref{Table_concept3}, and the linear optics functions are plotted in Figure~\ref{Figure_Twiss_concept3}.

\begin{table}[!h]
	\centering
	\caption{Parameters of the decay ring design with non-superconducting arc dipoles}
	\begin{tabular}{|p{0.25\textwidth}|p{0.2\textwidth}|}
		\hline
		Parameters & Values (units)\\
		\hline
		\hline
		Circumference & 535.9 (m)\\
		Arc length & 86.39 (m)\\
		Straight length & 181.56 (m)\\
		($\nu_x$, $\nu_y$) & (6.23, 7.21)\\
		($d\nu_x$/$d\delta$, $d\nu_y$/$d\delta$) & (-3.11,-12.73)\\
		\hline
	\end{tabular}
	\label{Table_concept3}
\end{table}

With respect to the design based on superconducting dipoles, the new design has a bigger circumference (536 compared to 488), and requires more combined function dipoles (24 compared to 13). The drift space between magnets is also too small for individuals sextupoles to be placed, therefore, the sextupole correction could only be added to the combined-function dipoles, making them multipoles, which still keeps the pole-tip field below 2 Tesla. The survival rate for 100 turns before sextupole corrections is 50\%.

\section{Sextupole correction optimization using a Genetic Algorithm}

Sextupole correction is needed in the ring to improve the acceptance of the ring. In this section, the correction scheme and its optimization for the ring design with superconducting combined function dipoles in the arcs are described.

Due to the space limits between magnets, sextupole fields also need to be added to the quadrupoles in the arcs to form combined function multipoles. Quadrupole-sextupole combined function magnets have been designed and tested by many experiments, for example in \cite{combined_function_magnet, BParker}. This arc design allows 5 individual sextupoles and 1 combined function multipole, all at dispersive locations, to carry sextupole fields.

The Genetic Algorithm \cite{GA_book} has been applied in many accelerator fields, including muon facilities \cite{NIMA_Ao_Horn}. The algorithm can test a wide range in parameter space to propose new generations of individual solutions that have a high probability of surpassing the older generation. In this study, the same GA module as described in \cite{NIMA_Ao_Horn}, written in Python, was used on NERSC (National Energy Research Scientific Computing Center). In this study, the single objective of the optimization is to maximize the number of particles that survive 100 turns. For tracking the MAD-X PTC TRACK module has been used. The 9 optimization variables include 6 sextupolar field strengths and 3 drift space lengths. The tracking is started from the middle of the production straight shown as the starting point in Figure~\ref{Figure_SC_DKring}. The initial muon beam is generated based on the covariance matrix of the beam at that point with the momentum of particles uniformly distributed within $3.8\pm10\%$ GeV/c, and the transverse phase space normally distributed within the 2000 $\mu$m$\cdot$rad acceptance. After optimization the acceptance, defined as the percentage of particles surviving 100 turns, was increased from 58\% to 67\%.

\begin{figure*}[htp]
	\centering
	\includegraphics[width=0.35\textwidth]{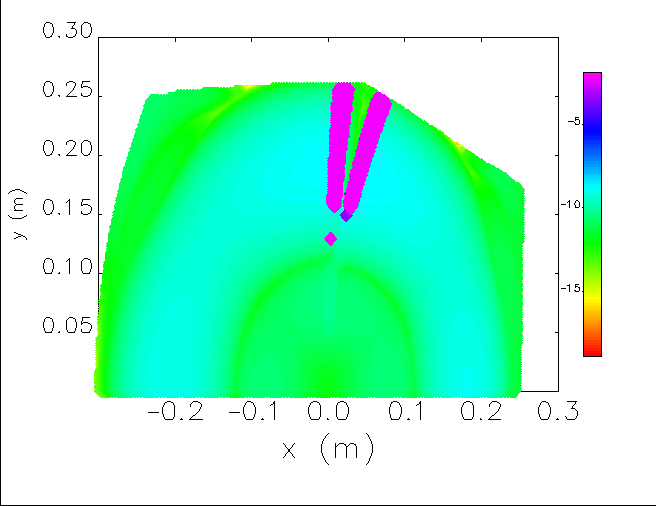}
	\includegraphics[width=0.35\textwidth]{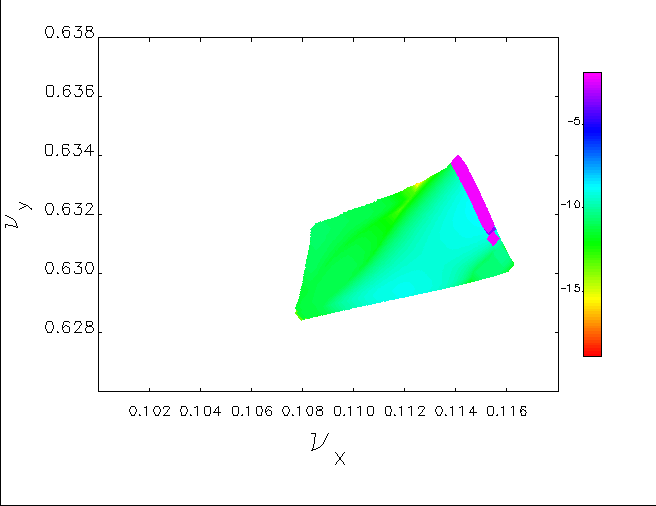}\\
	\includegraphics[width=0.35\textwidth]{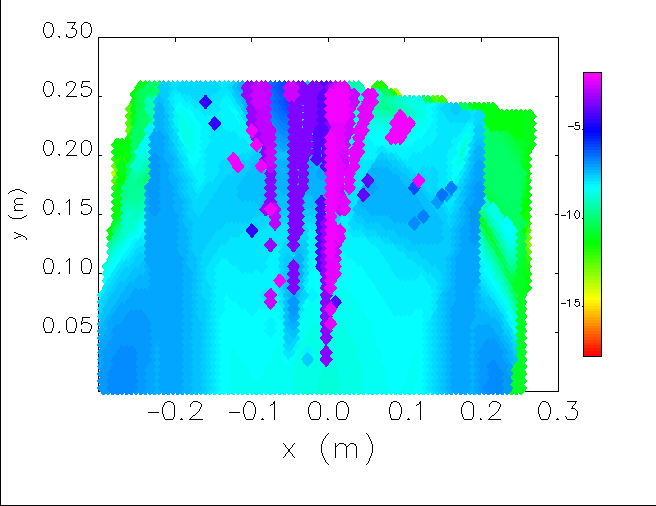}
	\includegraphics[width=0.35\textwidth]{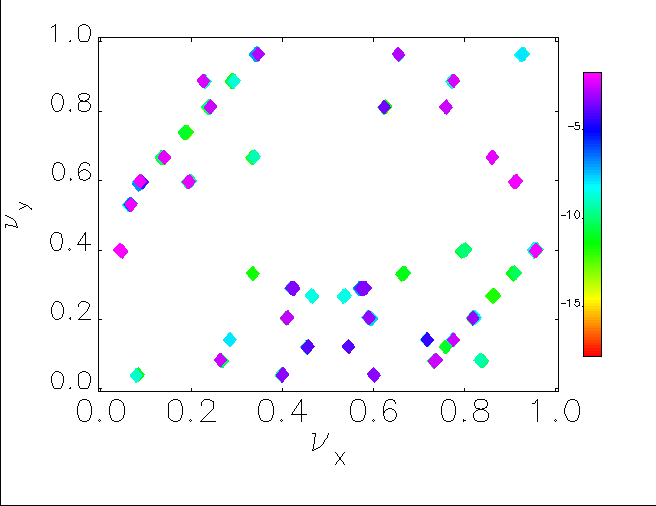}
	\caption{The FMA of the post-sextupole optimization ring design with combined function dipoles in the arcs in the X-Y space (left) and on the betatron tune plane (right), on-momentum particles (upper) and off-momentum particles within the full range of $\delta=\pm0.1$ (lower) are analyzed.}
	\label{Figure_FMA_concept1_661}
	\clearpage
\end{figure*}

The FMA of the post-optimization ring design is shown in Figure~\ref{Figure_FMA_concept1_661}. The FMA results imply that the sextupole corrections are not capable of correcting all the higher order terms.

\begin{figure}[htp]
	\centering
	\includegraphics[width=0.495\textwidth]{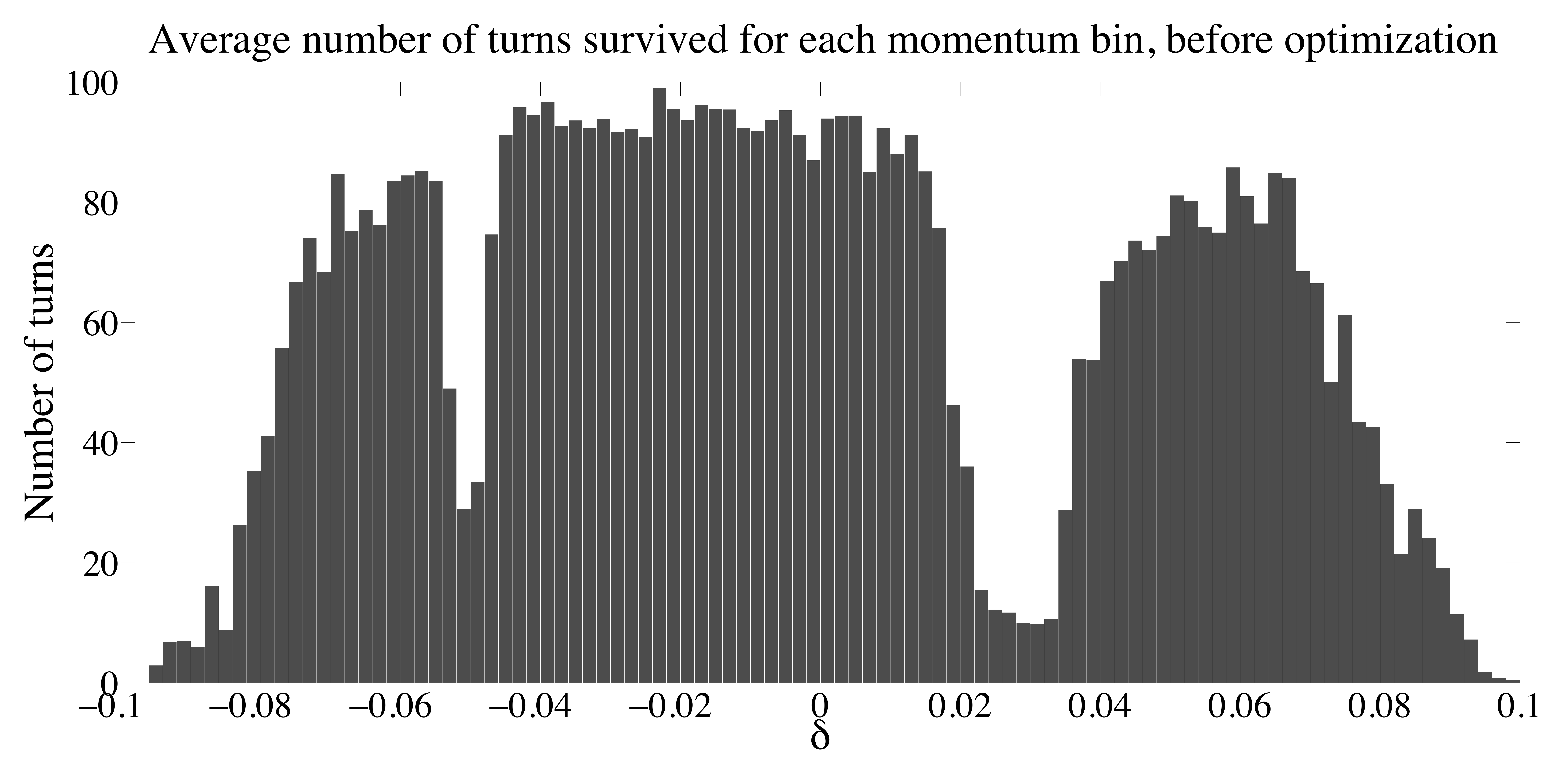}
	\includegraphics[width=0.495\textwidth]{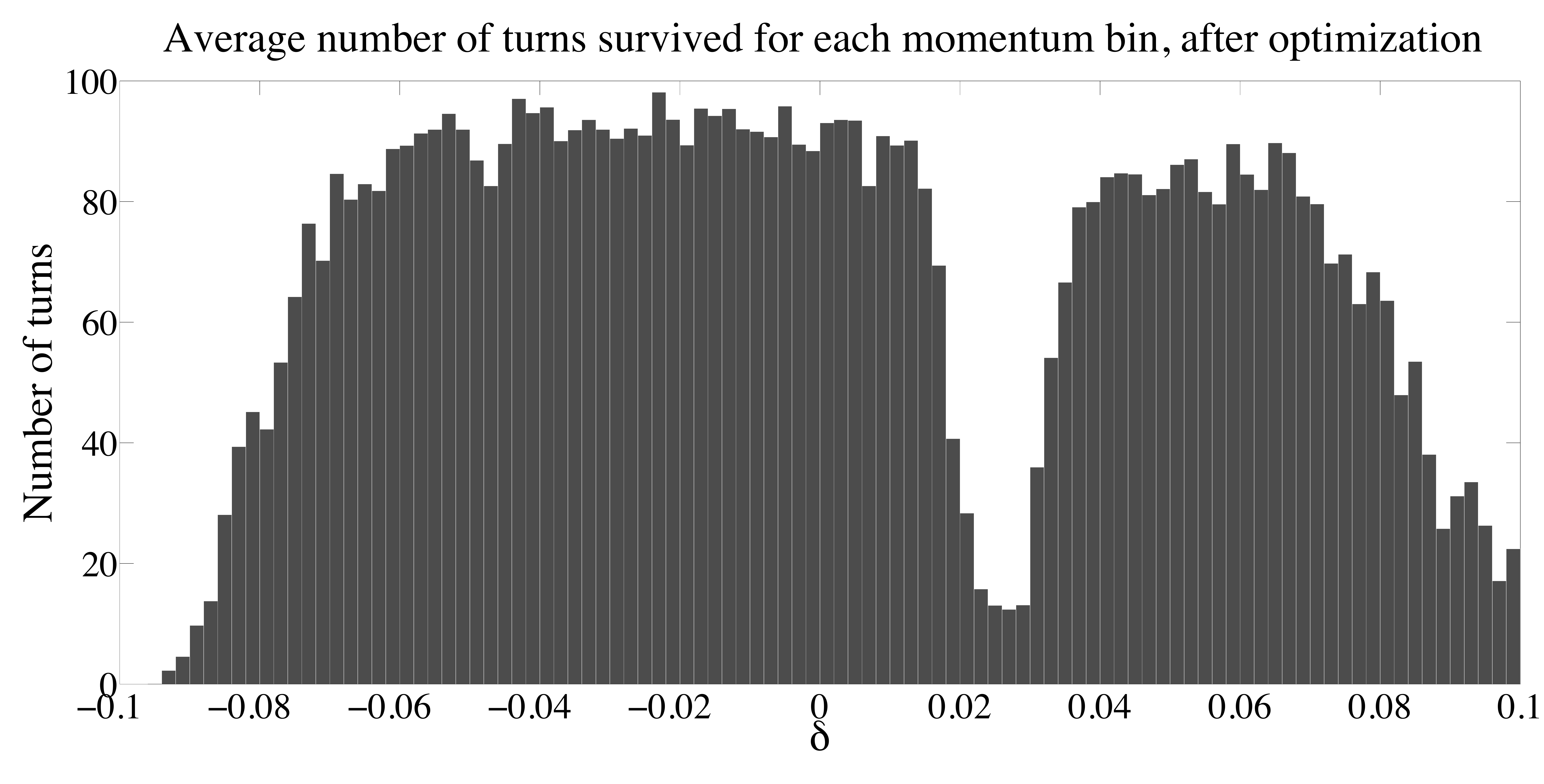}
	\caption{The average number of turns that particles in each momentum bin survive. The initial beam has a 300 $\mu$m$\cdot$rad transverse RMS emittance and a momentum spread of $\delta\in\pm10\%$. The upper and lower plots correspond to data from the pre-optimized lattice and post-optimized lattice, respectively.}	
	\label{Figure_avrg_turns_momentum_1_1}
\end{figure}

The histograms of the average survival turns for particles in each momentum bin are shown in Figure~\ref{Figure_avrg_turns_momentum_1_1}. Both the pre-optimization and post-optimization lattice tracking results are plotted and compared.
It is observed that, there are major stopbands at two $\delta$ locations, $\delta\approx-0.05$ and $\delta\approx0.03$. When $\left|\delta\right|$ is close to 0.1, the acceptance is further limited by higher-order chromatic effects, such as the terms in Table~\ref{Table_chrom_optim}. The algorithm, while not able to correct all the nonlinear effects, finds the global optimum and nicely corrects one stopband.
\begin{table}[htp]
	\centering
	\caption{Comparison of the principal terms corrected by the sextupole optimization}
	\begin{tabular}{|p{0.1\textwidth}|p{0.15\textwidth}|p{0.15\textwidth}|}
		\hline
		Parameters & Before sextupole correction & After sextupole optimization\\
		\hline
		\hline
		$d\nu_x/d\delta$ &-4.29 & -4.61\\
		$d^2\nu_x/d\delta^2$ &-3.62 & 0.41\\
		$d^3\nu_x/d\delta^3$ &-326 & 166\\
		$d\nu_y/d\delta$ & -7.34 & -6.68\\
		$d^2\nu_y/d\delta^2$ &-10.7 & -3.79\\
		$d^3\nu_y/d\delta^3$ &-94.1 & -78.5\\
		$D_2$ & 21.2 & 1.57\\
		$D_3$ & 958 & 831\\
		\hline
	\end{tabular}
	\label{Table_chrom_optim}
\end{table}

\begin{figure}[htp]
	\centering
	\includegraphics[width=0.495\textwidth]{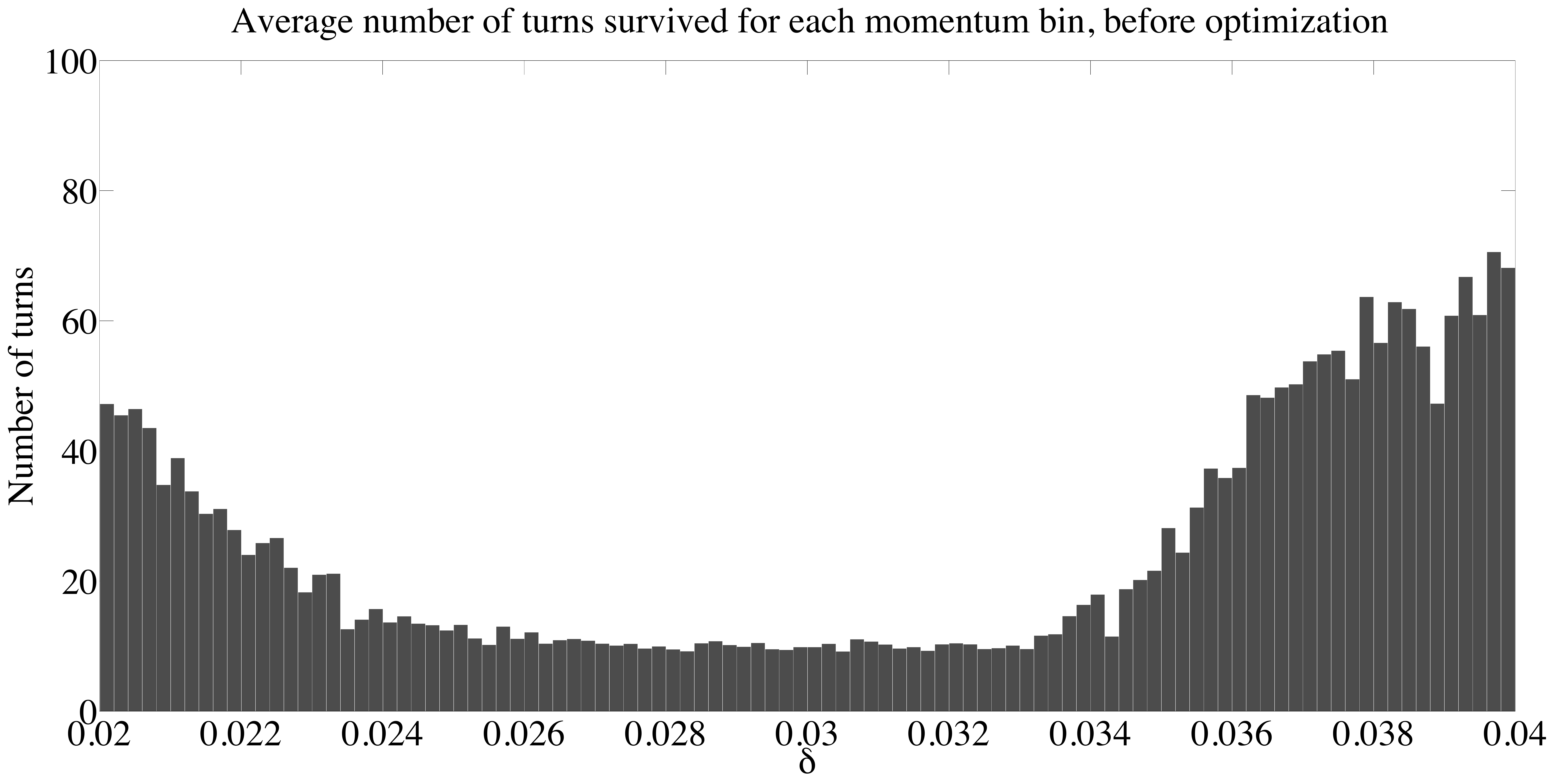}
	\includegraphics[width=0.495\textwidth]{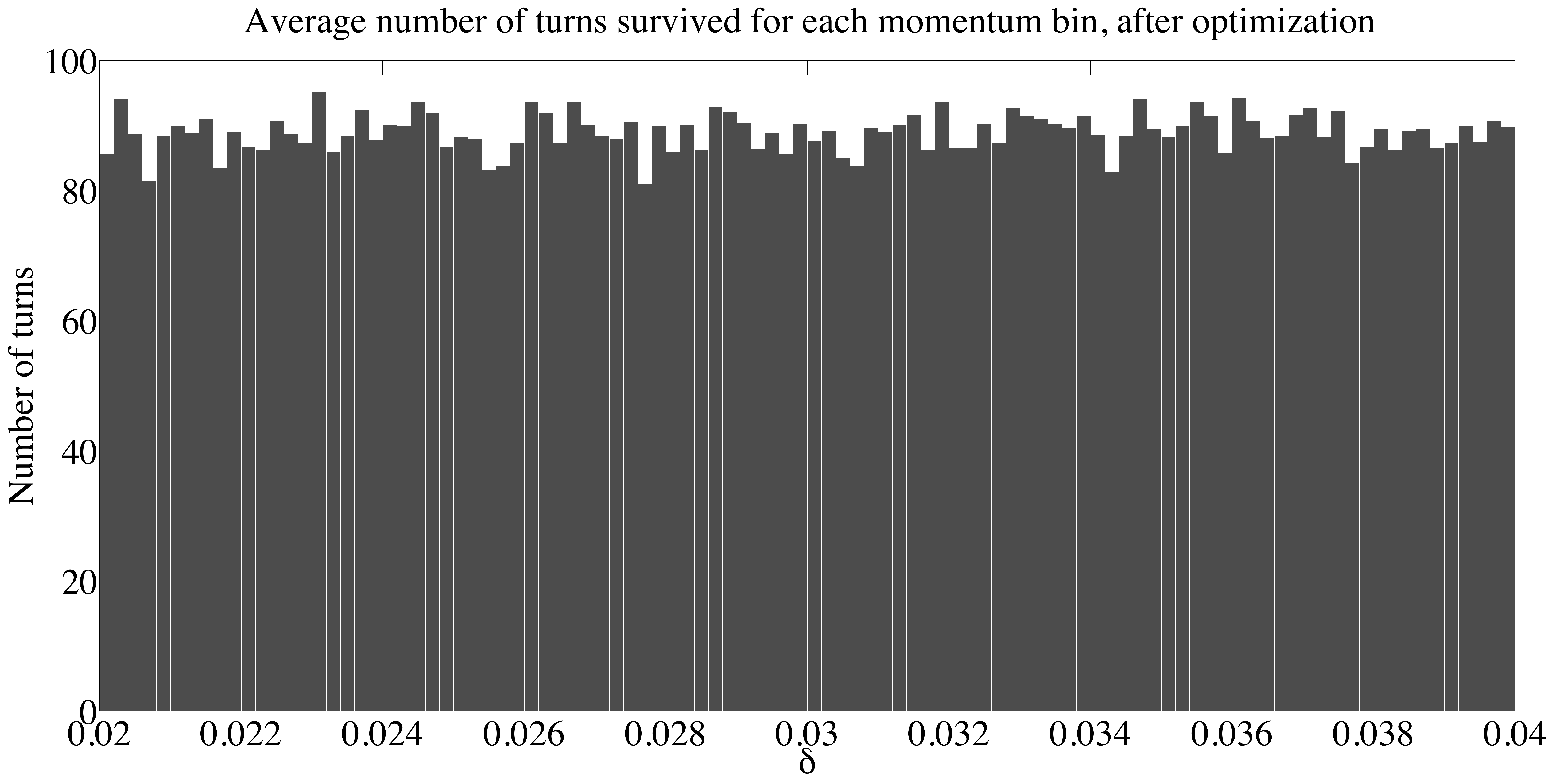}
	\caption{The average number of turns that particles in each momentum bin survive. The data were obtained from tracking of a beam with 300 $\mu$m$\cdot$rad transverse emittance and a momentum spread of $\delta\in[0.02,0.04]$, which is a limited band that is not improved by the full optimization in Figure~\ref{Figure_avrg_turns_momentum_1_1}. The upper and lower plots correspond to data from the pre-optimized and post-optimized lattice, respectively.}	
	\label{Figure_avrg_turns_momentum_1_5}
\end{figure}

In order to validate the optimization and investigate on the $\delta\sim0.03$ stopband shown in Figure~\ref{Figure_avrg_turns_momentum_1_1} that the GA did not improve, a special test beam, with 300 $\mu$m$\cdot$rad transverse RMS emittance and $\delta\in[0.02,0.04]$ was generated for tracking. The optimization was run again to find the optimum correction scheme specifically for this beam. The acceptance for this beam was increased from only 16\% to 88\% by the algorithm. As a comparison, the average number of survived turns for particles in each momentum bin, corresponding to Figure~\ref{Figure_avrg_turns_momentum_1_1}, is plotted in Figure~\ref{Figure_avrg_turns_momentum_1_5}.

\begin{figure}[htp]
	\centering
	\includegraphics[width=0.495\textwidth]{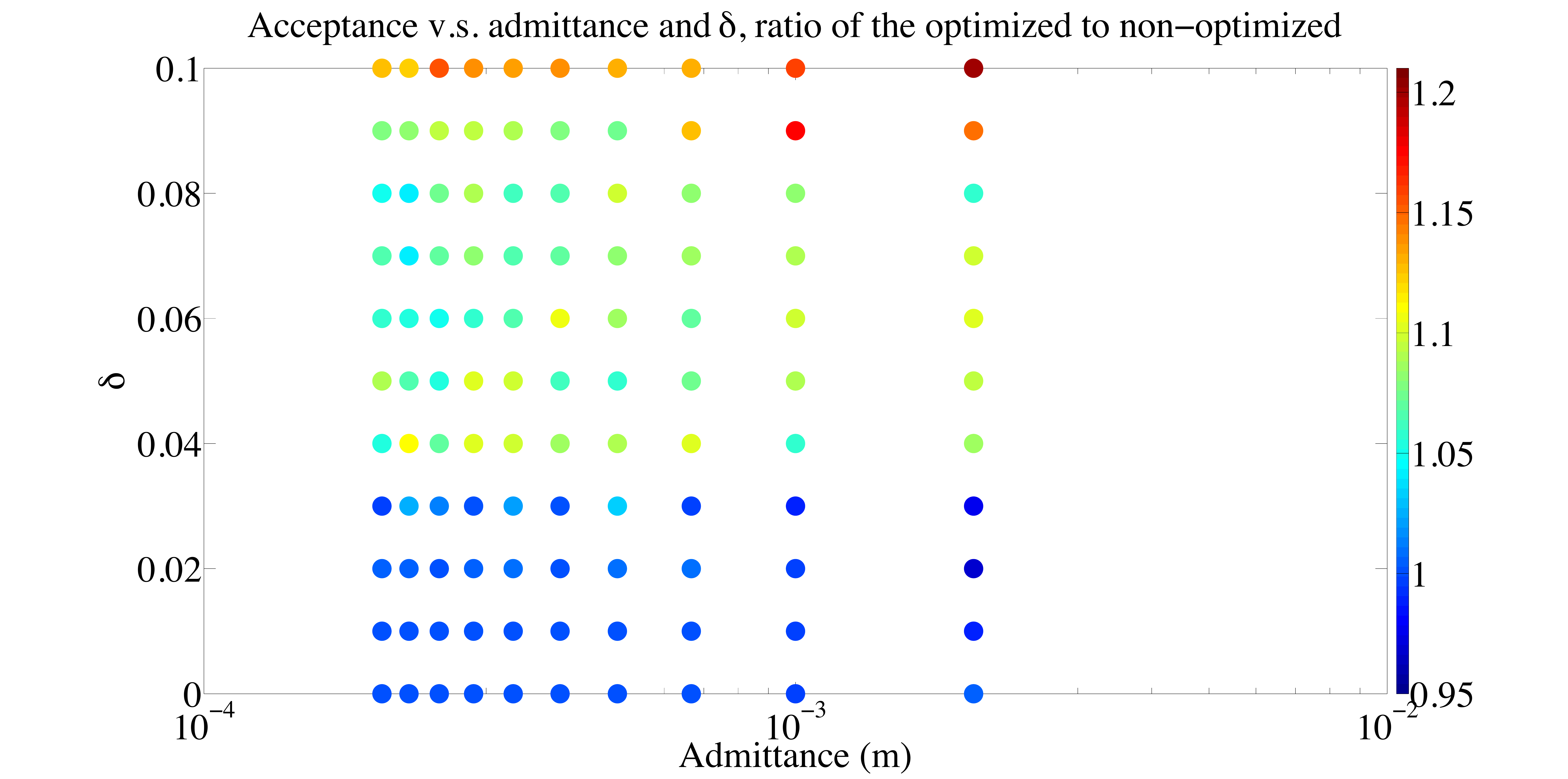}
	\includegraphics[width=0.495\textwidth]{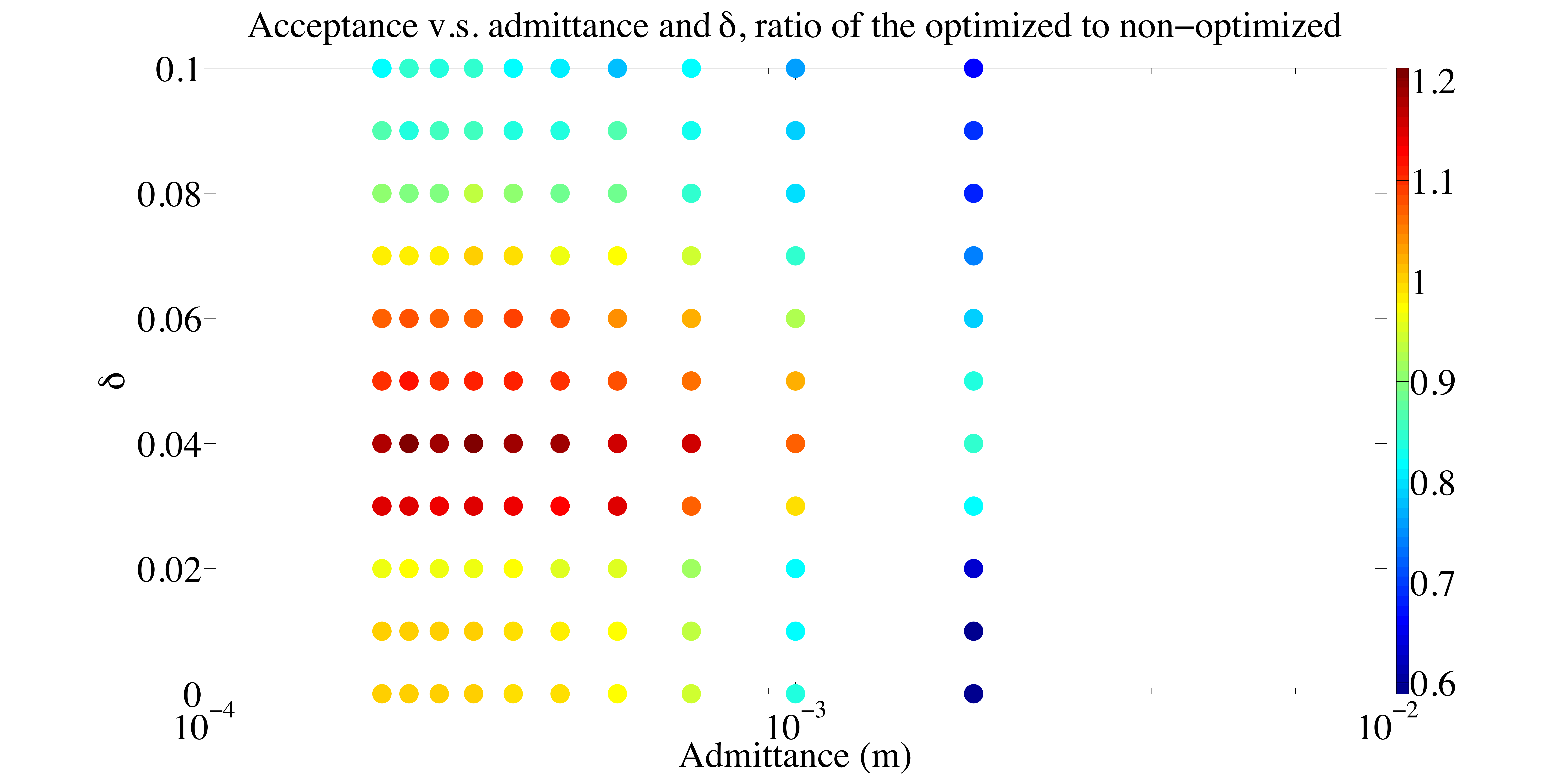}
	\caption{The increase in acceptance of different beams, represented by the ratio of the acceptance of the optimized lattices to the linear lattice. Upper: the original optimized lattice; Lower: the optimized lattice for a beam with 300 $\mu$m$\cdot$rad transverse admittance and a momentum spread of $\delta\in[0.02,0.04]$.}	
	\label{Figure_runCompare_ratio_1_5}
\end{figure}

It can be seen that, the stopband at $\delta=0.03$ was successfully corrected by the optimization. The comparison of this correction scheme and the optimized correction scheme is shown in Figure~\ref{Figure_runCompare_ratio_1_5}. The color of each marker represents the increase in the acceptance at each $\delta-\epsilon$ value combination where $\epsilon$ is the emittance of the beam. The low acceptance of the full-sized beam (upper-right points of the lower picture in Figure~\ref{Figure_runCompare_ratio_1_5} supports the hypothesis that the correction of this momentum stopband creates other nonlinearities and thus is not a globally optimum configuration. The figure indicates that for finding a global optimum the optimization must be carried out for the actual beam.

\section{Conclusions}
The nuSTORM muon decay ring aims to accept a muon beam with both a large transverse admittance of 2000 $\mu m$ and a large momentum range of $3.8\pm10\%$ GeV/c. It has been shown that the beam is greatly affected by the nonlinearities of the lattice. Due to the constraints on the ratio of arc section lengths to straight section lengths, the sextupole correction for the nonlinearities is limited by the number of sextupoles or sextupolar fields that can be placed in the ring and their strengths. A design with arc cells formed by superconducting combined function dipoles was discussed in this paper. From a multi-particle tracking in which the beam was normally distributed with a 300 $\mu$m$\cdot$rad RMS emittance and uniformly distributed in momentum within $3.8\pm10\%$ GeV/c, 58\% of the particles circulate 100 turns. The sextupole correction scheme was optimized with a genetic algorithm, the acceptance of the designs is increased to 67\%. The effectiveness of the optimization was validated by using different initial beams and optimizations. It was shown that the heuristic algorithms can find the best balance among all the dominant nonlinearities for each beam, and find the global optimum of the problem.

\section{acknowledgment}
The authors thank Dr. Mark Palmer for his supports on nuSTORM, Dr. David Adey for the discussions on the neutrino physics, and the MAP program colleagues on the discussions about the design. This research used resources of the National Energy Research Scientific Computing Center, a DOE Office of Science User Facility supported by the Office of Science of the U.S. Department of Energy under Contract No. DE-AC02-05CH11231.

\bibliographystyle{ieeetr}
\bibliography{nuSTORM_ring}

\begin{thebibliography}{10}

\bibitem{nuSTORM_Proposal}
P.~Kyberd {\em et~al.}, ``{nuSTORM - Neutrinos from STORed Muons: Letter of
  Intent to the Fermilab Physics Advisory Committee},'' 2012.

\bibitem{nuSTORM_PRD}
D.~Adey {\em et~al.}, ``Light sterile neutrino sensitivity at the nustorm
  facility,'' {\em Phys. Rev. D}, vol.~89, p.~071301, Apr 2014.

\bibitem{NeufferStochastic}
D.~Neuffer, ``Design considerations for a muon storage ring,'' 1980.
\newblock Telmark Conference on Neutrino Mass, Barger and Cline eds., Telmark,
  Wisconsin.

\bibitem{Ao_iPAC2013}
D.~Neuffer and A.~Liu, ``Stochastic injection scenarios and performance for
  nustorm,'' {\em Proc. IPAC2013, Shanghai, China, TUPFI055}, p.~1469, 2013.

\bibitem{Ao_iPAC2013_talk}
A.~Liu, A.~Bross, D.~Neuffer, and S.~Lee, ``$\nu$ storm facility design and
  simulation,'' 2013.

\bibitem{NIMA_pion_beamline}
A.~Liu {\em et~al.}, ``Design and simulation of the nustorm pion beamline,''
  {\em Nuclear Instruments and Methods in Physics Research Section A},
  vol.~801, pp.~44--50, 2015.

\bibitem{G4BL_website}
T.~Robert, ``G4beamline main website.''
  \url{http://http://www.muonsinternal.com/muons3/G4beamline}.

\bibitem{NIMA_Ao_Horn}
A.~Liu {\em et~al.}, ``Optimization of the magnetic horn for the nustorm
  non-conventional neutrino beam using the genetic algorithm,'' {\em Nuclear
  Instruments and Methods in Physics Research Section A}, vol.~794,
  pp.~200--205, 2015.

\bibitem{MARS15}
N.~Mokhov, ``Mars15, website link
  http://www-ap.fnal.gov/mars/intro\_manual.html.''

\bibitem{SY_book}
S.~Lee, {\em Accelerator Physics}.
\newblock World Scientific, 3rd~ed., 2012.

\bibitem{elegant_MB}
M.~Borland, ``Advanced photon source, anl.''
  \url{http://www.aps.anl.gov/Accelerator_Systems_Division/Accelerator_Operations_Physics/software.shtml#elegant}.

\bibitem{MADX}
L.~Deniau {\em et~al.}, ``Mad project main web page.''
  \url{http://mad.web.cern.ch/mad/}, 2002-Now.

\bibitem{MADX_PTC_Schmidt}
F.~Schmidt, ``Mad-x ptc integration,'' in {\em Particle Accelerator Conference,
  2005. PAC 2005. Proceedings of the}, pp.~1272--1274, May 2005.

\bibitem{combined_function_magnet}
W.~Beeckman {\em et~al.}, ``Design, fabrication, measurement, installation and
  alignment of 2 types of quadrupole-sextupole combined magnets for the upgrade
  of the 1.2 gev booster synchrotron at tohoku university,'' in {\em
  Proceedings of the Particle Accelerator Conference, 2013. PAC 2013.},
  pp.~1229--1231, 2013.

\bibitem{BParker}
B.~Parker, ``Superconducting magnets for use inside the hera ep interaction
  regions,'' in {\em Epac98}, vol.~TUOA02A, 1998.

\bibitem{GA_book}
M.~Mitchell, {\em An Introduction to Genetic Algorithms}.
\newblock MIT Press Cambridge, 1998.

\end{thebibliography}
\clearpage

\end{document}